\documentclass[11pt,a4paper]{article}
\usepackage{jheppub}
\usepackage[english]{babel}
\usepackage{amssymb,amsmath}
\usepackage{hyperref}
\usepackage{color}
\usepackage{graphicx}
\usepackage{mathtools}
\usepackage{subfigure}
\usepackage{slashed}
\usepackage{float}
\usepackage{booktabs}

\renewcommand{\Im}{\operatorname{Im}}

\title{The Continuum Dark Matter Zoo}

\author[a]{Csaba Cs\'aki,}
\author[a]{Ameen Ismail,}
\author[b]{and Seung J. Lee}
 
\affiliation[a]{Laboratory for Elementary Particle Physics, Cornell University, Ithaca, NY 14853, USA}
\affiliation[b]{Department of Physics, Korea University, Seoul 136-713, Korea}

\emailAdd{csaki@cornell.edu}
\emailAdd{ai279@cornell.edu}
\emailAdd{sjjlee@korea.ac.kr}

\abstract{
    We generalize the recently proposed continuum dark matter model to the case where the dark matter consists of a spin-$1/2$ or spin-$1$ gapped continuum. We construct simple continuum analogs of weakly interacting massive particles annihilating through the $Z$ portal. We discuss all existing experimental constraints, with the strongest bounds arising from indirect detection and limits on continuum decays from the cosmic microwave background. Our models are phenomenologically viable for gap scales of $60$--$200$~GeV (spin-$1/2$) and $35$--$90$~GeV (spin-$1$), owing to the strong kinematic suppression of direct detection bounds which is unique to continuum states. We comment on future prospects for detection and suggest directions for further continuum model building.
}

\begin{document}

\maketitle	
\flushbottom

%%%%%%%%%%%%%%%%%%%%%%%%%%%%%%%%%%%%%%%%%%%%%%%%%%%%%%%%%%%
%%%%%%%%%%%%%%%%%%%%%%%%%%%%%%%%%%%%%%%%%%%%%%%%%%%%%%%%%%%
\section{Introduction}
%%%%%%%%%%%%%%%%%%%%%%%%%%%%%%%%%%%%%%%%%%%%%%%%%%%%%%%%%%%
%%%%%%%%%%%%%%%%%%%%%%%%%%%%%%%%%%%%%%%%%%%%%%%%%%%%%%%%%%%

The existence of dark matter (DM) is supported by a plethora of experimental observations and provides robust evidence for new physics beyond the Standard Model (SM). Yet the microscopic nature of DM remains elusive, due to the lack of observation of nongravitational signatures. New physics models for DM span a mass range of about 80 orders of magnitude. On the lower end, ultralight bosonic DM may be as light as $\sim 10^{-22}$~eV, with the lower bound arising from the de Broglie wavelength of the DM becoming comparable to galactic scales~\cite{Hu:2000ke}. On the upper end, primordial black holes can be the primary component of DM for masses less than $\sim 10^{48}$~GeV, with an upper bound derived from the nonobservation of microlensing~\cite{Villanueva-Domingo:2021spv}. 

Historically much attention has been focused on weakly interacting massive particle (WIMP) candidates, which have electroweak-scale masses and annihilation cross sections, due to theoretical motivation from Higgs naturalness and the ``WIMP miracle''. By now, however, the simplest models of WIMPs annihilating to SM particles through the $Z$ or Higgs portal are mostly excluded by direct detection experiments. This has spurred intense exploration of DM models beyond the WIMP paradigm, and also motivated the development of more sophisticated WIMP models that evade direct detection bounds.

Here we consider the possibility that dark matter consists not of ordinary particles, but rather of a continuum of states. The unique kinematics of continuum DM has distinctive phenomenological consequences, including a strong suppression of direct detection bounds~\cite{Csaki:2021gfm,Csaki:2021xpy}.

\begin{figure}
    \centering
    \includegraphics[width=0.8\textwidth]{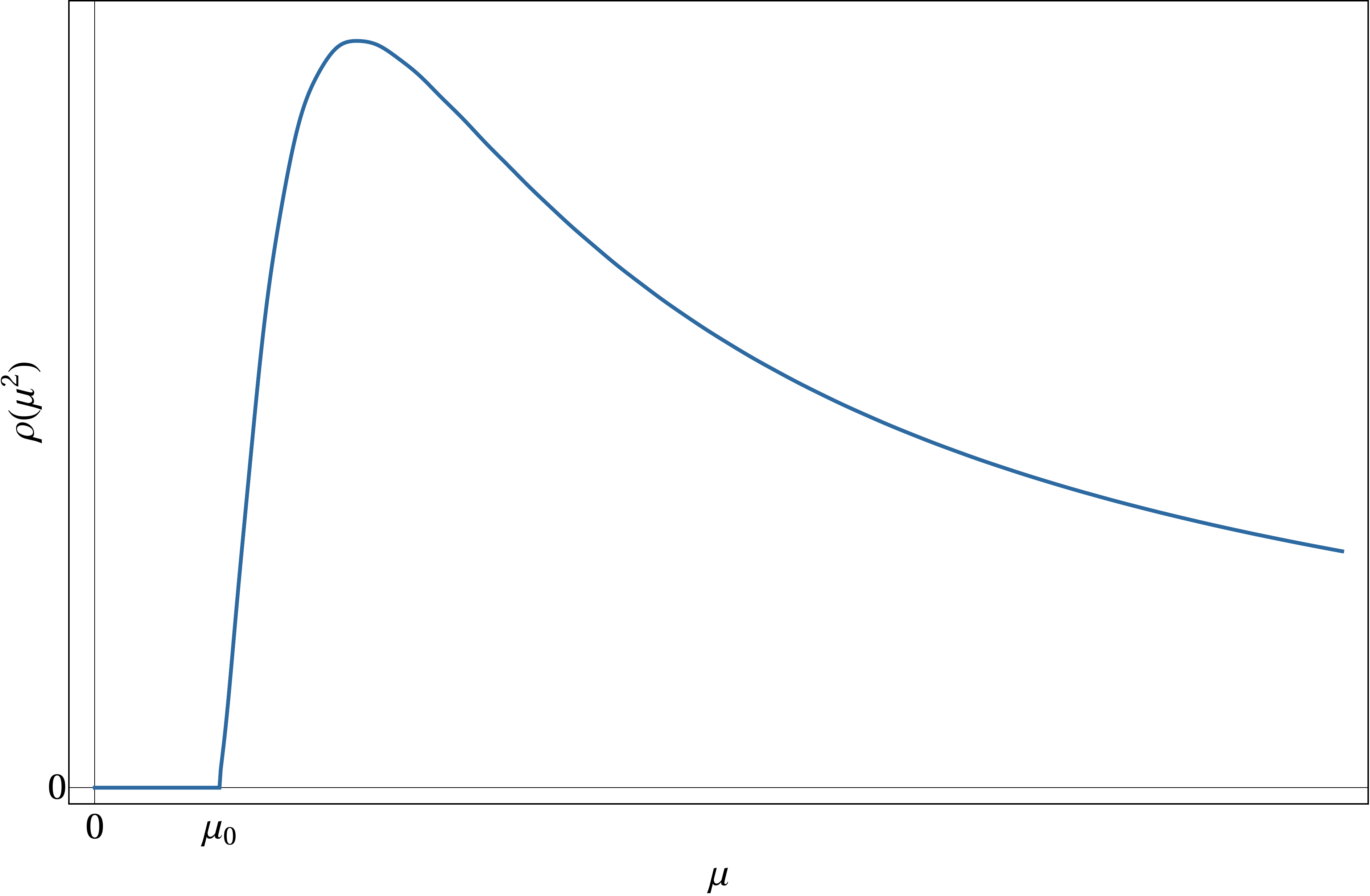}
    \caption{Cartoon of a gapped continuum spectral density. The spectral density vanishes for $\mu$ less than the gap scale $\mu_0$.}
    \label{fig:rhocartoon}
\end{figure}

A continuum field is characterized by its spectral density $\rho(\mu^2)$. For an ordinary particle of mass $m$, the spectral density is just $2\pi \delta(\mu^2 - m^2)$, but for a continuum field it is a more complicated smooth function. We are interested specifically in gapped continua, where $\rho(\mu^2)$ vanishes for $\mu^2 < \mu_0^2$, where $\mu_0^2$ is the gap scale. An example spectral density is depicted in fig.~\ref{fig:rhocartoon}.

Continuum states and related ideas have seen many applications in model building, since continuous spectra occur generically in scale-invariant theories. These include unparticles~\cite{Georgi:2007ek,Georgi:2007si}, approaches to Higgs naturalness based on compositeness or extra dimensions~\cite{Falkowski:2008fz,Stancato:2008mp,Falkowski:2008yr,Bellazzini:2015cgj,Csaki:2018kxb,Cabrer:2009we,Megias:2019vdb,Megias:2021mgj}, and continuum supersymmetry models~\cite{Cai:2009ax,Cai:2011ww,Gao:2019gfw}. In the context of DM, the past few years have seen the development of models with continuum states mediating between a dark sector and the SM~\cite{Katz:2015zba,Chaffey:2021tmj}, as well as DM from a hidden conformal sector~\cite{Hong:2019nwd,Hong:2022gzo,Chiu:2022bni}.

Refs.~\cite{Csaki:2021gfm,Csaki:2021xpy} introduced the study of DM models based on a gapped continuum, focusing on a simple model of scalar continuum DM. Similar to a WIMP, the gap scale of the continuum is around the electroweak scale and the DM interacts with the Standard Model via the $Z$ portal. Importantly, if the DM were an ordinary particle in this model, it would have been excluded long ago by direct detection experiments. But the kinematics of continuum states suppresses direct detection bounds, and hence the scalar $Z$-portal continuum DM model is viable phenomenologically. This is not the only distinctive consequence of continuum kinematics. Any DM state with $\mu > \mu_0$ is necessarily unstable because it can decay to a lighter DM state. Late-time decays can inject electromagnetic energy into the SM sector after recombination, reionizing hydrogen; consequently, observations of the cosmic microwave background (CMB) strongly constrain continuum DM. Continuum states also produce novel signatures at colliders, as they undergo a cascade of decays to progressively lighter states.

Note that there is a somewhat similar framework called dynamical dark matter (DDM)~\cite{Dienes:2011ja,Dienes:2011sa}, which has been explored in a series of articles by Dienes, Thomas and collaborators~\cite{Dienes:2012yz,Dienes:2012cf,Dienes:2013xya,Dienes:2014via,Dienes:2014bka,Boddy:2016fds,Curtin:2018ees,Dienes:2019krh}. The key idea of DDM is that the dark matter consists of a tower of long-lived particles, resulting in multi-component dark matter. While being different in theoretical nature, continuum DM models share certain phenomenological properties with DDM. For example, one can observe a similar modification of the direct detection cross sections and energy spectra~\cite{Dienes:2012cf,Dienes:2013xya}, exotic collider signatures~\cite{Dienes:2012yz,Dienes:2014bka,Dienes:2019krh}, and decays among the constituents of the dark sector leading to an evolving DM distribution~\cite{Dienes:2014via,Curtin:2018ees}. However, a lot of essential characteristics of continuum DM models are quite unique and distinguishable from generic properties of DDM~\cite{Csaki:2021gfm}.

In this paper we explore continuum DM models where the DM is a spin-$1/2$ fermion or spin-$1$ vector, rather than a scalar. The striking effects of continuum kinematics --- particularly the suppression of direct detection and the existence of late-time decays --- are still present as in the case of scalar continuum DM. We continue to focus on DM which annihilates through the $Z$ portal. As in refs.~\cite{Csaki:2021gfm,Csaki:2021xpy}, the simple models we construct would have been ruled out by direct detection experiments for ordinary particle DM. However, continuum DM easily satisfies all experimental constraints.

This paper is organized as follows. We first review basic aspects of continuum states and applications to DM models in sec.~\ref{sec:continuum}. (We prove some results about the form of the spectral density near the gap scale for fermions and vector bosons in appendix~\ref{sec:appendix}.) This provides the necessary foundation for our models of fermion and vector $Z$-portal continuum DM, which we describe in secs.~\ref{sec:fermionmodel} and~\ref{sec:vectormodel}, respectively. We find that both models can reproduce the correct thermal relic density while satisfying all experimental constraints. The fermion model is phenomenologically viable for gap scales in the range $60$~GeV~to~$200$~GeV, and the vector model for gap scales in the range $35$~GeV~to~$90$~GeV. Lastly, we conclude and outline future avenues for continuum model building in sec.~\ref{sec:conclusions}.

%%%%%%%%%%%%%%%%%%%%%%%%%%%%%%%%%%%%%%%%%%%%%%%%%%%%%%%%%%%%%%%%%%%
%%%%%%%%%%%%%%%%%%%%%%%%%%%%%%%%%%%%%%%%%%%%%%%%%%%%%%%%%%%%%%%%%%%
%%%%%%%%%%%%%%%%%%%%%%%%%%%%%%%%%%%%%%%%%%%%%%%%%%%%%%%%%%%%%%%%%%%
%%%%%%%%%%%%%%%%%%%%%%%%%%%%%%%%%%%%%%%%%%%%%%%%%%%%%%%%%%%%%%%%%%%

%%%%%%%%%%%%%%%%%%%%%%%%%%%%%%%%%%%%%%%%%%%%%%%%%%%%%%%%%%%
%%%%%%%%%%%%%%%%%%%%%%%%%%%%%%%%%%%%%%%%%%%%%%%%%%%%%%%%%%%
\section{Continuum physics review}\label{sec:continuum}
%%%%%%%%%%%%%%%%%%%%%%%%%%%%%%%%%%%%%%%%%%%%%%%%%%%%%%%%%%%
%%%%%%%%%%%%%%%%%%%%%%%%%%%%%%%%%%%%%%%%%%%%%%%%%%%%%%%%%%%
We begin with a whirlwind tour of the essentials of continuum physics. For a more thorough investigation, the reader is referred to refs.~\cite{Csaki:2021gfm,Cabrer:2009we,Megias:2019vdb}.

\subsection{Continuum fields}

The quadratic action for a scalar continuum field can be written in momentum space as
\begin{equation}
    S_{\rm scalar} = \int \frac{d^4 p}{(2\pi)^4} \phi^\dagger(p) \Sigma(p^2) \phi(p)
\end{equation}
where the function $\Sigma$ is determined by the two-point correlation function. It is related to the spectral density by
\begin{equation}\label{eq:scalarspectraldensity}
    \rho(\mu^2) = -2 \Im \frac{1}{\Sigma(\mu^2)}  \quad \Leftrightarrow \quad \frac{1}{\Sigma(p^2)} = \int \frac{d \mu^2}{2\pi} \frac{\rho(\mu^2)}{p^2-\mu^2+i\epsilon} .
\end{equation}
An ordinary particle of mass $m$ is described by $\Sigma(p^2) = p^2 - m^2 + i \epsilon$, leading to $\rho(\mu^2) = 2\pi\delta(\mu^2 - m^2)$. Recall that for a gapped continuum, $\rho(\mu^2)$ vanishes for $\mu < \mu_0$ and is described by some smooth function for $\mu > \mu_0$.

The situation for fermions and vector bosons is similar. The quadratic actions are
\begin{equation}\begin{split}
    S_{\rm fermion} &= -i \int \frac{d^4 p}{(2\pi)^4} \overline{\psi}(p) \Sigma(p^2) \frac{\overline{\sigma}^\mu p_\mu}{p^2} \psi(p) \\
    S_{\rm vector} & = \frac{1}{2} \int \frac{d^4 p}{(2\pi)^4} A_\mu(p) \Sigma(p^2) \left[ \eta^{\mu\nu} - (1 - \frac{1}{\xi}) \right] A_\nu(p)
\end{split}\end{equation}
where $\psi$ is a left-handed Weyl fermion, $A_\mu$ is a vector field, and $\xi$ is the usual gauge-fixing parameter. The spectral density is again given in terms of $\Sigma$ by eq.~\eqref{eq:scalarspectraldensity}.

The details of the spectral density function depend on its origin. A gapped continuum can arise as a 4D effective description of a 5D model based on a warped extra dimension~\cite{Cabrer:2009we,Megias:2019vdb}. The spectral density can be numerically computed once the 5D geometry and parameters in the bulk Lagrangian (such as a bulk mass term) are specified.
Upon choosing a particular 5D background (like the ``soft wall'' background of ref.~\cite{Cabrer:2009we}), we find the spectral density evaluated near the gap scale $\mu_0$ always takes the form
\begin{equation}\label{eq:universalspectraldensity}
    \rho(\mu^2) = \frac{\rho_0}{\mu_0^2} \sqrt{\frac{\mu^2}{\mu_0^2} -1}
\end{equation}
where $\rho_0$ is a constant which depends on the Lagrangian. Ref.~\cite{Csaki:2021gfm} proved this for scalar fields. In appendix~\ref{sec:appendix}, we review the derivation of spectral densities from warped constructions, then prove that fermion and vector spectral densities also take this universal form near the gap scale.

In the remainder of this paper we will not be concerned with the microscopic origin of the gapped continuum. We will only assume the spectral density obeys eq.~\eqref{eq:universalspectraldensity} near the gap scale. When necessary to choose a particular value of $\rho_0$ to compute an experimental bound, we will take $\rho_0 = 2\pi$, but we will be explicit when doing so.

\subsection{Interactions}
One can add interaction terms coupling continuum fields to other fields just as for ordinary particles. For example, in sec.~\ref{sec:fermionmodel} we will have a continuum fermion $\psi$ interacting with the $Z$ boson through a coupling $\overline{\psi} \slashed{Z} \psi$. We assume the couplings are sufficiently weak that they do not impact the form of the spectral density.

Computing decay widths and cross sections with continuum states proceeds as for ordinary particles, except that we must integrate over the spectral density. That is, the phase space integral for a continuum state is
\begin{equation}
    \int \frac{d\mu^2}{2\pi} \rho(\mu^2) \int \frac{d^3 \vec{p}}{(2\pi)^3} \frac{1}{2\sqrt{\mu^2 + \lvert \vec{p} \rvert^2}} .
\end{equation}

In calculating the relic abundance of continuum DM, one finds that the Boltzmann equation takes the same form as for particle DM. All of the continuum effects are captured by the thermally averaged annihilation cross section $\langle \sigma v \rangle$. The computation of $\langle \sigma v \rangle$ involves integrals over spectral densities, but because of Boltzmann factors $e^{-\mu/T}$, these integrals are dominated by values of $\mu$ close to $\mu_0$. For this reason, the annihilation cross section of continuum DM is typically the same as for particle DM of mass $\mu_0$, up to continuum corrections of order $T/\mu_0$\footnote{More rigorously, one evaluates the spectral density integrals using a saddle-point approximation, assuming the universal form of the spectral density $\rho \propto \sqrt{(\mu/\mu_0)^2-1}$ near the gap scale.}. Assuming the DM freezes out around $\mu_0/T = 10$--$30$, we expect the size of these corrections to be $\lesssim 10\%$.

We caution that it is only valid to approximate the continuum DM as a particle of mass $\mu_0$ here because of the Boltzmann factors in the thermally averaged cross section. In general, continuum effects cannot be ignored and one must carefully evaluate or approximate the spectral density integrals. This leads to the unique effects mentioned earlier: late-time decays of DM and suppression of direct detection rates. In the next two sections, we will see these effects explicitly in the phenomenology of our fermion and vector continuum models.

%%%%%%%%%%%%%%%%%%%%%%%%%%%%%%%%%%%%%%%%%%%%%%%%%%%%%%%%%%%
%%%%%%%%%%%%%%%%%%%%%%%%%%%%%%%%%%%%%%%%%%%%%%%%%%%%%%%%%%%
\section{Spin-$1/2$ model}\label{sec:fermionmodel}
%%%%%%%%%%%%%%%%%%%%%%%%%%%%%%%%%%%%%%%%%%%%%%%%%%%%%%%%%%%
%%%%%%%%%%%%%%%%%%%%%%%%%%%%%%%%%%%%%%%%%%%%%%%%%%%%%%%%%%%
To build a model of continuum fermion DM, we add to the SM a singlet continuum left-handed fermion $\psi_L$ and a mediator Dirac fermion $\chi$, which is an SU(2) doublet with hypercharge $Y = 1/2$.
The most general effective 4D Lagrangian is then
\begin{equation}
    \mathcal{L} = \mathcal{L}_{\rm SM} + \mathcal{L}_\psi + \mathcal{L}_\chi + \mathcal{L}_{\rm int} ,
\end{equation}
where
\begin{equation}\begin{split}
    \mathcal{L}_\psi &= i \overline{\psi}_L \Sigma (p^2) \overline{\sigma}^\mu p_\mu \psi_L \\
    \mathcal{L}_\chi &= i \overline{\chi}_L \overline{\sigma}^\mu D_\mu \chi_L + i \overline{\chi}_R \sigma^\mu D_\mu \chi_R - M (\overline{\chi}_L \chi_R + \overline{\chi}_R \chi_L) \\
    \mathcal{L}_{\rm int} &= -\kappa \overline{\chi}_R \psi_L  H .
\end{split}\end{equation}
As explained in sec.~\ref{sec:continuum}, the continuum spectrum encoded by $\Sigma(p^2)$ can arise from a bulk fermion in a warped extra dimension.

Assuming the Dirac mass $M$ of $\chi$ is much larger than the gap scale $\mu_0$, we can integrate out its neutral component $\chi^0$. The classical equations of motion for $\chi^0$ yield 
\begin{equation}
    \chi_L^0 = - \frac{\kappa v M}{\sqrt{2}(\partial^2 + M^2)} \psi_L, \quad \chi_R^0 = - i \frac{\kappa v }{\sqrt{2}(\partial^2 + M^2)} \overline{\sigma}^\mu \partial_\mu \psi_L .
\end{equation}
Substituting this solution back into the action yields an effective Lagrangian for $\psi_L$. We find that $\psi_L$ acquires an effective coupling to the $Z$ of the form
\begin{equation}
    -\frac{g_Z}{2} \left[ \frac{\kappa v M}{\sqrt{2}(M^2 - \mu^2)} \right]^2 \overline{\psi}_L \slashed{Z} \psi_L
\end{equation}
where $g_Z = g/\cos \theta_w = \sqrt{g^2 + g'^2}$. This essentially describes $\mu$-dependent mixing between $\psi_L$ and $\chi_L^0$. In the limit $\mu \ll M$, the $\mu$-dependence is negligible, and the effective interaction can be written in terms of a mixing angle
\begin{equation}
    \sin \alpha = \frac{\kappa v}{\sqrt{2}M} .
\end{equation}

Additionally, $t$-channel exchange of a $\chi_R^0$ leads to a dimension-five interaction between the continuum state and the Higgs,
\begin{equation}
    \frac{\kappa^2}{M} H^\dagger H \overline{\psi}_L \psi_L = \frac{\sqrt{2} \kappa \sin \alpha}{v} H^\dagger H \overline{\psi}_L \psi_L .
\end{equation}
This interaction induces exotic Higgs decays, $h \rightarrow \overline{\psi}_L \psi_L$, which imposes a constraint on the model for $\mu_0 < m_h/2$. Otherwise, the Higgs coupling does not introduce any qualitative change to the phenomenology, so we neglect it in our freeze-out calculations.

In summary the effective theory below the Dirac mass $M$ contains a continuum left-handed fermion with a coupling to the $Z$:
\begin{equation}
    \mathcal{L}_{\rm eff} = \mathcal{L}_{\rm SM} + \mathcal{L}_\psi - \frac{g_Z \sin^2 \alpha}{2} \overline{\psi}_L \slashed{Z} \psi_L .
\end{equation}
We have described one way, but certainly not the only way, of generating this model. For the remainder of this section we will work in the context of the effective theory.
We caution that experimental bounds require the Dirac fermion to be heavier than $300$--$500$~GeV, depending on its branching fractions~\cite{Falkowski:2013jya}; this implies an upper bound on the mixing angle of $\sin \alpha \lesssim 0.3$--$0.6$. In some parts of the parameter space we require larger values of $\sin \alpha$ to achieve the observed DM relic abundance, so a different UV completion is needed.

\subsection{Freeze-out}
The computation of the relic abundance of fermionic continuum DM proceeds similarly to ordinary particle Majorana $Z$-portal DM. For $\mu_0 < m_W$, the only relevant process for DM annihilation is $\psi\overline{\psi} \rightarrow f\overline{f}$ via s-channel $Z$ exchange, where $f$ is any SM fermion other than the top quark. (We will drop the subscript on $\psi_L$ for clarity.) The thermally averaged cross section is
\begin{equation}
    \left\langle \sigma v (\psi\overline{\psi} \rightarrow f\overline{f}) \right\rangle = \frac{ g_Z^2 \sin^4 \alpha}{32 \mu_0^2} \frac{\Gamma_Z}{m_Z} \left[ \left(1 - \frac{m_Z^2}{4\mu_0^2}\right)^2 + \frac{m_Z^2 \Gamma_Z^2}{16\mu_0^4}\right]^{-1} v_{\rm rel}^2 ,
\end{equation}
where $v_{\rm rel}$ is the relative velocity and $\Gamma_Z$ is the $Z$ decay width. The $v_{\rm rel}^2$ suppression indicates this process is p-wave.
When $\mu_0 > m_W$, the DM can also annihilate to $W^+ W^-$ through s-channel $Z$ exchange\footnote{
For $\mu_0 \sim m_W$, the three-body decay $\psi\overline{\psi} \rightarrow W W^* \rightarrow W \ell \nu$ may be important, and so the thermally averaged annihilation cross section may require an $\mathcal{O}(1)$ correction in the immediate vicinity of the $W$ mass. This is also the case for scalar and vector $Z$-portal DM, and it does not qualitatively affect our conclusions.
}, leading to a cross section
\begin{equation}
    \left\langle \sigma v (\psi\overline{\psi} \rightarrow W^+ W^-) \right\rangle = \frac{ g_Z^4 \sin^4 \alpha}{96 \pi \mu_0^2} \left(1 - \frac{1}{y}\right)^{3/2} \left( \frac{4y^2 + 20y + 3}{y^2}\right) \left( \frac{x}{4 - 1/x} \right)^2 v_{\rm rel}^2
\end{equation}
where $x = \mu_0^2 / m_Z^2$ and $y = \mu_0^2/m_W^2$. We see this is also p-wave.

For $\mu_0 > m_Z$, the process $\psi\overline{\psi} \rightarrow ZZ$ becomes possible, mediated by $t$-channel $\psi$ exchange. The cross section is suppressed by an extra factor of $\sin^4 \alpha$ relative to the $f\overline{f}$ and $WW$ channels. It is not negligible though, because we require $\sin^4 \alpha \gtrsim 0.1$ to achieve the correct relic abundance. To compute the amplitude for the process, one must evaluate an integral over the spectral density for the internal $\psi$ propagator, which takes the form
\begin{equation}\label{eq:UVdependence}
    \int \frac{d\mu^2}{2\pi} \frac{\rho(\mu^2)}{t - \mu^2 }  = c_{\rm UV} \times  \frac{\rho_0/2\pi}{t - \mu_0^2}.
\end{equation}
This integral depends upon the behaviour of $\rho(\mu^2)$ at large $\mu \gg \mu_0$, and therefore it depends on the UV completion of the gapped continuum. We parametrize this by an unknown, order-one coefficient $c_{\rm UV}$. Fortunately, we will see momentarily the strongest experimental constraints for $\mu_0 > m_Z$ come from indirect detection bounds, which directly constrain $\langle \sigma v\rangle$. Thus, the dependence on the UV completion essentially cancels out (i.e. the value of $\sin^2 \alpha$ which yields the correct relic abundance and the indirect detection limit on $\sin^ 2 \alpha$ depend on $c_{\rm UV}$ in the same way). We can make conclusions regarding the experimental viability of the model which are independent of the UV completion.

For concreteness, we will take $c_{\rm UV} = 1$ and $\rho_0 = 2\pi$ for relic abundance calculations. We then find
\begin{equation}
    \left\langle \sigma v (\psi\overline{\psi} \rightarrow ZZ) \right\rangle = \frac{g_Z^4 \sin^8 \alpha}{512\pi \mu_0^2} \left( 1 - \frac{1}{x} \right)^{3/2} \left( \frac{4x^2 + 20x + 5}{x^2} \right) \left( \frac{x}{2 - 1/x} \right)^2 .
\end{equation}
This process is s-wave.

Finally, annihilations to $Zh$ and $t \overline{t}$ are also mediated by $s$-channel $Z$-exchange when $\mu_0 > (m_Z + m_h)/2$ and $\mu_0 > m_t$, respectively. The thermally averaged cross sections are
\begin{equation}\begin{split}
    \left\langle \sigma v (\psi\overline{\psi} \rightarrow Zh) \right\rangle &= \frac{g_Z^4 \sin^4 \alpha}{4096 \pi \mu_0^4 m_Z^4} \left[ 16\mu_0^4 - 8\mu_0^2 (m_Z^2+m_h^2) + (m_Z^2-m_h^2)^2 \right]^{3/2} , \\
    \left\langle \sigma v (\psi\overline{\psi} \rightarrow t\overline{t}) \right\rangle &=  \frac{3 g_Z^4 \sin^4 \alpha m_t^2}{128 \pi m_Z^4} \sqrt{1 - \frac{m_t^2}{\mu_0^2}}
\end{split}\end{equation}
which are also s-wave.

\begin{figure}
    \centering
    \includegraphics[width=0.8\textwidth]{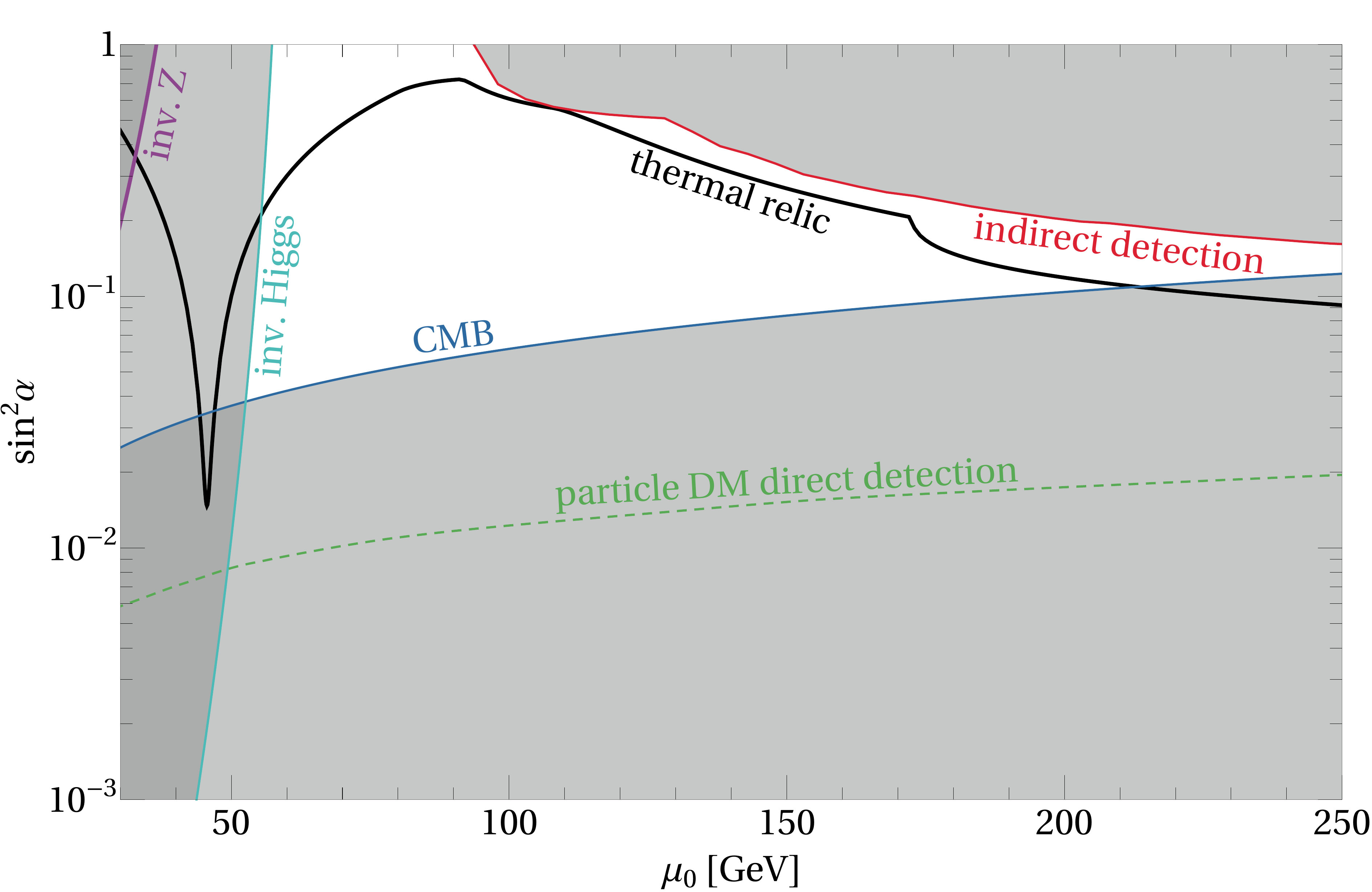}
    \caption{Parameter space for continuum fermion $Z$-portal DM. The black curve reproduces the observed thermal relic density~\cite{Planck:2015fie}. We show experimental constraints from CMB observations \textit{(blue)}~\cite{Slatyer:2016qyl,Slatyer:2015jla}, indirect detection with the Fermi-LAT experiment \textit{(red)}~\cite{Fermi-LAT:2016uux,Abazajian:2020tww,Arcadi:2014lta}, the LEP bound on the invisible $Z$ width \textit{(purple)}~\cite{ALEPH:2005ab}, and the LHC limit on the invisible Higgs branching ratio \textit{(turquoise)}~\cite{CMS:2022qva,ATLAS:2022yvh}. We also show the LUX-ZEPLIN direct detection bound for ordinary particle Majorana $Z$-portal DM \textit{(green, dashed)}~\cite{LZ:2022ufs}.}
    \label{fig:fermionbounds}
\end{figure}

In fig.~\ref{fig:fermionbounds} we show the region of parameter space that yields the observed dark matter relic density, alongside current bounds from indirect detection, the CMB, and collider experiments. Continuum fermion $Z$-portal DM is consistent with all experimental constraints for gap scales $\mu_0 \in (60, 200)$~GeV. We also show direct detection limits for the analogous model of particle DM in fig.~\ref{fig:fermionbounds}. Direct detection rules out ordinary particle $Z$-portal spin-$1/2$ DM, except for the case of a purely axial vector coupling and a very large mass~\cite{Arcadi:2017kky,Ellis:2017ndg,LZ:2022ufs}. Continuum kinematics makes direct detection bounds weaker for our model by three to four orders of magnitude, so that they do not even show up on our plot. We will return to this point shortly --- we first discuss the other bounds in detail.

%%%%%%%%%%%%%%%%%%%%%%%%%%%%%%%%%%%%%%%%%%%%%%%%%%%%%%%%%%%%%%%%%%%
%%%%%%%%%%%%%%%%%%%%%%%%%%%%%%%%%%%%%%%%%%%%%%%%%%%%%%%%%%%%%%%%%%%
%%%%%%%%%%%%%%%%%%%%%%%%%%%%%%%%%%%%%%%%%%%%%%%%%%%%%%%%%%%%%%%%%%%
%%%%%%%%%%%%%%%%%%%%%%%%%%%%%%%%%%%%%%%%%%%%%%%%%%%%%%%%%%%%%%%%%%%

\subsection{Experimental bounds}

\subsubsection{Indirect detection}

Indirect detection is irrelevant for $\mu_0 < m_Z$, because of the $v^2$ suppression of the annihilation cross section. For $m_Z < \mu_0 < (m_Z + m_h)/2$ the strongest indirect detection constraint comes from the Fermi-LAT upper bound on the $ZZ$ annihilation cross section~\cite{Fermi-LAT:2016uux,Abazajian:2020tww}. For $\mu_0 > (m_Z + m_h)/2$ we also consider the Fermi bound on the $Zh$ cross section computed in Ref.~\cite{Arcadi:2014lta}. In fig.~\ref{fig:fermionbounds} we plot the stronger of these two constraints.

\subsubsection{CMB}

The remaining bounds are not independent of $\rho_0$, so we will take $\rho_0 = 2\pi$. The CMB bound arises from electromagnetic energy injection by late decays $\psi(\mu) \rightarrow \psi(\mu') e^+ e^-$. Such decays can reionize hydrogen after recombination, leading to tension with CMB anisotropy observations~\cite{Slatyer:2016qyl,Slatyer:2015jla}. We require that decays to $e^+ e^-$ are kinematically forbidden by the time of CMB decoupling, which yields a lower limit on the mixing angle $\sin \alpha$.

More specifically, the rate of the $\psi(\mu) \rightarrow \psi(\mu') f\overline{f}$ decay, at leading order in $\Delta\mu = \mu - \mu'$, is given by
\begin{equation}\label{eq:latedecaywidth}
    \frac{8\sqrt{2}\rho_0}{45045\pi^4} \times \frac{g_Z^4 \sin^4 \alpha \left( c_A^2 + c_V^2 \right) \mu_0^5}{m_Z^4} \times \left( \frac{\Delta\mu}{\mu_0} \right)^{13/2}
\end{equation}
where $c_{A,V}$ are the usual axial vector and vector couplings of the fermion to the $Z$. One can estimate the maximum $\Delta \mu$ by equating this decay width (summed over $e^+ e^-$ and $\nu \overline{\nu}$ channels) to the Hubble parameter at the time of CMB decoupling, because heavier states would have already decayed before decoupling. (Ref.~\cite{Csaki:2021xpy} confirmed that this approximation agrees well with detailed numerical studies of the time evolution of the continuum mass distribution.) We impose $\Delta \mu < 2 m_e$ to obtain the resulting bound on the parameter space. Note that eq.~\eqref{eq:latedecaywidth} is identical to the corresponding decay width for scalar continuum $Z$-portal DM, up to a factor of $3/2$~\cite{Csaki:2021xpy}. For this reason the CMB bound in fig.~\ref{fig:fermionbounds} is just the same as that in the scalar case, rescaled by a factor of $\sqrt{2/3}$.

Late-time DM decays may also cause photodissociation of nuclei, altering the light element abundances predicted by Big Bang nucleosynthesis (BBN)~\cite{Reno:1987qw,Kawasaki:1994sc,Cyburt:2002uv,Kawasaki:2004qu,Hisano:2009rc}, but the resulting bound on the parameter space is weaker than the CMB anisotropy bound by about an order of magnitude~\cite{Henning:2012rm,Csaki:2021xpy}. Current bounds on DM decay from CMB spectral distortions are even weaker than the BBN bound~\cite{Chluba:2011hw,Chluba:2013wsa,Poulin:2016anj,Lucca:2019rxf}. Future CMB experiments such as PIXIE would tighten bounds on spectral distortions by about three orders of magnitude, potentially yielding a constraint competitive with the CMB anisotropy bound~\cite{Lucca:2019rxf,Kogut:2011xw,Chluba:2019nxa}.

\subsubsection{Invisible decays}

For $\mu_0 < m_Z / 2$, the decay $Z \rightarrow \psi\overline{\psi}$ is kinematically possible, and precision measurements of the $Z$ decay width constrain the parameter space. The decay width is given by
\begin{equation}\label{eq:fermionZdecay}\begin{split}
    \left( \frac{\rho_0}{2\pi} \right)^2 &\frac{g_Z^2 \sin^4 \alpha}{96 \pi} m_Z \int d(t_1^2) \int d(t_2^2) \sqrt{t_1^2-1} \sqrt{t_2^2-1} \\
    &\times \sqrt{1 - 2x(t_1^2+t_2^2) + x^2(t_1^2-t_2^2)^2} \left[2 - x(t_1^2+6t_1t_2+t_2^2) - x^2(t_1^2-t_2^2)^2\right]
\end{split}\end{equation}
where $x = \mu_0^2 / m_Z^2$, and the integration region is defined by $t_1 > 1$, $t_2 > 1$, and $y_1 + y_2 < 1/\sqrt{x}$. (These are just integrals over the spectral density where we have rescaled $\mu \rightarrow \mu/\mu_0$.) We compare this decay width to the LEP bound on new invisible $Z$ decays assuming lepton flavor universality, $\Gamma_{\rm inv} < 2$~MeV~\cite{ALEPH:2005ab}. This excludes the model for $\mu_0 \lesssim 33$~GeV.

Recall that the dimension-five term
\begin{equation}
    \frac{\sqrt{2} \kappa \sin \alpha}{v} H^\dagger H \overline{\psi}_L \psi_L
\end{equation}
leads to exotic Higgs decays $h \rightarrow \psi\overline{\psi}$ for $\mu_0 < m_h/2$. The decay width is
\begin{equation}\begin{split}
    \left( \frac{\rho_0}{2\pi} \right)^2 &\frac{\kappa^2 \sin^2 \alpha}{4 \pi} m_h \int d(t_1^2) \int d(t_2^2) \sqrt{t_1^2-1} \sqrt{t_2^2-1} \\
    &\times \sqrt{1 - 2x'(t_1^2+t_2^2) + x'^2(t_1^2-t_2^2)^2} \left[1 - x'(t_1 + t_2)^2 \right]
\end{split}\end{equation}
where $x' = \mu_0^2 / m_h^2$, and $t_1$, $t_2$, and the integration region are defined as in eq.~\eqref{eq:fermionZdecay}. The Higgs invisible branching fraction is constrained to be less than $0.15$~\cite{CMS:2022qva,ATLAS:2022yvh}. We plot the resulting bound in fig.~\ref{fig:fermionbounds}, taking $\kappa=1$ for simplicity.

\subsubsection{Direct detection}

Let us return now to direct detection. We consider the spin-dependent cross section for DM-nucleon scattering $\psi(\mu) N \rightarrow \psi(\mu') N$; the spin-independent cross section is suppressed by a factor of $v^2$, just like ordinary particle Majorana $Z$-portal DM. In the nonrelativistic limit the spin-dependent cross section is
\begin{equation}\label{eq:DDcrosssection}
    \int \frac{d \mu'^2}{2\pi} \rho(\mu'^2) \frac{3 g_Z^4 \sin^4 \alpha \mu_{\psi N}^2 f_N^2}{4 \pi m_Z^4} ,
\end{equation}
where $\mu_{\psi N}$ is the reduced mass of the DM and the nucleon and $f_N$ is the axial vector-like $Z$-nucleon coupling. ($f_N$ is determined by the $Z$ axial coupling to quarks, $c_A/2$, and nuclear form factors, which we obtain from ref.~\cite{Hill:2014yxa}.) The integral is taken over all kinematically accessible masses for the outgoing DM state, ranging from $\mu_0$ to $\mu_0 + \Delta \mu$. The range $\Delta \mu$ is given by
\begin{equation}\label{eq:DDkinrange}
    \Delta \mu = (\mu - \mu_0) + qv - \frac{q^2}{2\mu_{\psi N}},
\end{equation}
where $q$ is the momentum transfer and $v$ the DM velocity. Note that the last two terms are bounded from above by $qv - q^2/(2 \mu_{\psi N}) \leq \mu_{\psi_N} v^2 / 2$; since the DM is nonrelativistic with $v \sim 10^{-3}$ this upper bound is $\mathcal{O}({\rm keV})$. We will see that this is subleading to the $\mu-\mu_0$ term in eq.~\eqref{eq:DDkinrange}.

We can find the typical size of $\mu$ by matching the continuum decay rate (eq.~\eqref{eq:latedecaywidth}) to the Hubble parameter today, just as we matched it to the Hubble parameter at decoupling for the CMB bound. This yields
\begin{equation}
    \mu - \mu_0 \approx 0.4 {\rm~MeV~} \frac{(\mu_0/100{\rm~GeV})^{3/13}}{(\sin^2 \alpha / 0.01)^{4/13}} .
\end{equation}
In the region of parameter space of relevance to us, this contribution to $\Delta \mu$ dominates over the other terms in eq.~\eqref{eq:DDkinrange}, so $\Delta \mu \approx \mu - \mu_0$.
The spectral density integral then contributes a factor of
\begin{equation}
    \frac{\rho_0}{2\pi \mu_0^2} \int_{\mu_0^2}^{(\mu_0+\Delta\mu)^2} d\mu^2 \sqrt{ \frac{\mu^2}{\mu_0^2} - 1 } \approx \frac{ 2 \times 10^{-8} }{(\sin^2 \alpha/0.01)^{6/13} (\mu_0/100{\rm~GeV})^{15/13}}
\end{equation}
which strongly suppresses direct detection rates! One can then recast bounds from direct detection experiments by comparing the DM-nucleon cross section to that for ordinary particle DM. Since the cross section is proportional to $\sin^4 \alpha$, direct detection bounds on $\sin^2 \alpha$ are suppressed by three to four orders of magnitude.

In fig.~\ref{fig:fermionbounds} we show the direct detection bound in the absence of the continuum suppression (i.e. for ordinary particle DM), obtained from the latest LUX-ZEPLIN limit on the DM-neutron spin-dependent cross section~\cite{LZ:2022ufs}. We see that this model would be ruled out, were it not for continuum effects. Continuum kinematics weakens the bound so that direct detection does not constrain our model at all.

\subsection{Nonrenormalizable terms and tuning}
We have not discussed nonrenormalizable operators other than the Higgs coupling, including the dimension-five operators $\overline{\psi}\psi W^+ W^-$ and $\overline{\psi}\psi Z^2$ as well as the dimension-six operator $\overline{\psi}\psi \overline{f}f$. One expects these terms to be generated at a cutoff scale $\Lambda$, which will induce corrections to the $WW$ and $ZZ$ ($f\overline{f}$) annihilation cross sections of order $T/\Lambda$ (order $(T/\Lambda)^2$).

Recall that we are already ignoring continuum effects of order $T/\mu_0 \lesssim \mathcal{O}(10\%)$. As long as $\Lambda > \mu_0$, which is necessary for the consistency of the model, the effects of nonrenormalizable terms will be parametrically smaller than these effects.

Lastly we consider the tuning of the spectral density normalization $\rho_0$. If the continuum arises from a warped extra dimension, then $\Lambda \sim R^{-1}$, and the consistency condition amounts to $R^{-1} \gtrsim \mu_0$ (where $R$ is the location of the UV brane). The 5D gravitational EFT breaks down around $R^{-1} \sim \mu_0$ as the UV brane approaches a curvature singularity. The natural scale of $\rho_0$ is $(\mu_0 R)^2$ (see appendix~\ref{sec:appendix}). Thus, obtaining $\mathcal{O}(1)$ values of $\rho_0$ while having a mild hierarchy $\mu_0 \sim 0.1 R^{-1}$ na\"ively requires a percent-level tuning. For continuum scalars and vector bosons there is no way to alleviate this tuning since the gap scale is determined purely by the 5D geometry. For fermions there could be less tuning, since the gap scale is determined by the vev of a bulk scalar and an associated Yukawa coupling~\cite{Megias:2019vdb}. However we would need a detailed 5D model to make a precise statement about the tuning.

%%%%%%%%%%%%%%%%%%%%%%%%%%%%%%%%%%%%%%%%%%%%%%%%%%%%%%%%%%%
%%%%%%%%%%%%%%%%%%%%%%%%%%%%%%%%%%%%%%%%%%%%%%%%%%%%%%%%%%%
\section{Spin-$1$ model}\label{sec:vectormodel}
%%%%%%%%%%%%%%%%%%%%%%%%%%%%%%%%%%%%%%%%%%%%%%%%%%%%%%%%%%%
%%%%%%%%%%%%%%%%%%%%%%%%%%%%%%%%%%%%%%%%%%%%%%%%%%%%%%%%%%%
Next we turn to vector continuum DM. We extend the SM gauge group by a dark U(1)$_D$; the quadratic Lagrangian for the dark gauge field is
\begin{equation}
    \mathcal{L}_V = \frac{1}{2} V_\mu(p) \Sigma(p^2) \left[ \eta^{\mu\nu} - (1 - \frac{1}{\xi}) \right] V_\nu(p) ,
\end{equation}
where $\Sigma(p^2)$ encodes a gapped continuum spectral density. If U(1)$_D$ is unbroken, the spectral density also necessarily has a pole at $p^2 = 0$ due to gauge invariance. For our purposes, we will assume the abelian Higgs mechanism or Stueckelberg mechanism lifts the mass of the would-be zero mode, so that the spectral density describes a continuum with gap scale $\mu_0$ plus an ordinary particle of mass $M$. We want the DM relic abundance to be dictated by the continuum physics, rather than the ordinary particle state, so we will futher assume $M \gg \mu_0$. Then we can practically ignore the mass-$M$ mode in our freeze-out calculations. In contrast, if we instead had $M < \mu_0$, that mode would be the lightest DM state, and the DM would freeze out like an ordinary particle of mass $M$. We are not interested in that situation.

Constructing a suitable $Z$-portal interaction is not trivial, unlike for scalar or fermion DM, due to the restrictions imposed by gauge invariance. We will consider the following ``generalized Chern--Simons'' interaction, which is commonly used in the literature for vector $Z$-portal DM~\cite{Arcadi:2017kky,Mambrini:2009ad}:
\begin{equation}\label{eq:generalizedCS}
    \mathcal{L}_{\rm int} = \frac{g_Z \lambda}{4} \epsilon^{\mu\nu\rho\sigma} V_\mu Z_\nu V_{\rho\sigma}
\end{equation}
where $V_{\rho\sigma} = \partial_\rho V_\sigma - \partial_\sigma V_\rho$ is the $V$ field strength tensor, and the $g_Z$ normalization factor is just for convenience.

Presumably the term in eq.~\eqref{eq:generalizedCS} would arise from integrating out heavy chiral fermions in the UV theory. We will not present a UV completion, but let us outline how this term can be embedded in a consistent effective theory, with a construction similar to the Green--Schwarz mechanism from string theory. Suppose one introduces four left-handed Weyl fermions with $({\rm U(1)}_D, {\rm U(1)}_Y)$ charges
\begin{equation}\label{eq:weylfermions}
    \psi_1(1,1), \quad \psi_2(-1,1), \quad \psi_3(0,-1), \quad \psi_4(0,-1) .
\end{equation}
Then all ${\rm U(1)}^3$ and ${\rm gravity}^2 {\rm U(1)}$ anomalies cancel except for the ${\rm U(1)}_D^2 {\rm U(1)}_Y$ anomaly. This anomaly can be cancelled by adding the (classically gauge-variant) term in eq.~\eqref{eq:generalizedCS} with the coefficient $\lambda = 2 g_D^2/(3\pi^2)$, where $g_D$ is the dark gauge coupling~\cite{Dror:2017nsg}
\footnote{The beta function for $\alpha_D = g_D^2/(4\pi)$ is $\beta(\alpha_D) = 5\alpha_D^2/(6\pi)$, assuming no matter charged under U(1)$_D$ other than the Weyl fermions in eq.~\eqref{eq:generalizedCS} and a dark Higgs with charge 1. Requiring that we do not hit a Landau pole below $5$~TeV leads to an upper bound $\lambda \lesssim 0.8$ at $100$~GeV. For larger $\lambda$ a more sophisticated model is needed.}
. The fermions can be given masses by coupling to a scalar field with ${\rm U(1)}_D$ charge $1$ which acquires a vev; this could be the same as the abelian Higgs field that lifts the mass of the zero mode of the dark gauge boson.
For more detailed discussion of generalized Chern--Simons terms we refer the reader to refs.~\cite{Mambrini:2009ad,Anastasopoulos:2006cz,Antoniadis:2009ze,Dudas:2009uq,Dudas:2013sia}.

Lastly, we must forbid kinetic mixing between the dark matter and the $Z$ or photon, because otherwise it would be unstable. To this end we impose dark charge conjugation symmetry, under which $V_\mu \rightarrow -V_\mu$. The effective Lagrangian for our $Z$-portal continuum vector DM model is then
\begin{equation}
    \mathcal{L}_{\rm eff} = \mathcal{L}_{\rm SM} + \mathcal{L}_V + \mathcal{L}_{\rm int} .
\end{equation}

\subsection{Freeze-out}
The annihilation cross sections for the $f\overline{f}$, $WW$, $Zh$, and $t\overline{t}$ final states are similar to the fermion model:
\begin{equation}\begin{split}
    \left\langle \sigma v (VV \rightarrow f\overline{f}) \right\rangle &= \frac{g_Z^2 \lambda^2}{288 \mu_0^2} \left( \frac{\Gamma_Z}{m_Z}\right) \left[ \left(1 - \frac{m_Z^2}{4\mu_0^2}\right)^2 + \frac{m_Z^2 \Gamma_Z^2}{16\mu_0^4}\right]^{-1} v_{\rm rel}^4 ,\\
    \left\langle \sigma v (VV \rightarrow W^+ W^-) \right\rangle &= \frac{g_Z^4 \lambda^2}{864 \pi \mu_0^2} \left(1 - \frac{1}{y}\right)^{3/2} \left( \frac{4y^2 + 20y + 3}{y^2}\right) \left( \frac{x}{4 - 1/x} \right)^2 v_{\rm rel}^4 ,\\
    \left\langle \sigma v (VV \rightarrow Zh) \right\rangle &= \frac{g_Z^4 \lambda^2}{36864 \pi \mu_0^4 m_Z^4} \left[ 16\mu_0^4 - 8\mu_0^2 (m_Z^2+m_h^2) + (m_Z^2-m_h^2)^2 \right]^{3/2} v_{\rm rel}^2 ,\\
    \left\langle \sigma v (VV \rightarrow t\overline{t}) \right\rangle &= \frac{g_Z^4 \lambda^2 m_t^2}{384 m_Z^4} \sqrt{1 - \frac{m_t^2}{\mu_0^2}} v_{\rm rel}^2
\end{split}\end{equation}
where $x = \mu_0^2/m_Z^2$ and $y = \mu_0^2/m_W^2$. A notable difference, however, is that the cross sections have an additional $v_{\rm rel}^2$ suppression relative to the fermion model. In particular the $f\overline{f}$ and $WW$ channels scale as $v_{\rm rel}^4$, indicating that these processes are d-wave. This same d-wave suppression is seen in the case of real scalar dark matter annihilating to light fermions~\cite{Kumar:2013iva,Toma:2013bka,Giacchino:2013bta}.

Like in the fermion model, the process $VV \rightarrow ZZ$ is mediated by $t$-channel DM exchange, and depends on the integral of the spectral density and thus on the UV completion. We will again parametrize our ignorance with a coefficient $c_{\rm UV}$ (see eq.~\eqref{eq:UVdependence}), then set $c_{\rm UV} = 1$ and $\rho_0 = 2\pi$ for the sake of concreteness. We then find
\begin{equation}
    \left\langle \sigma v (VV \rightarrow ZZ) \right\rangle = \frac{g_Z^4 \lambda^4}{4608\pi \mu_0^2} \sqrt{1 - \frac{1}{x}} \left[ \frac{x^4 + 12x^3 +88x^2 - 8x + 3}{(1 - 2x)^2} \right].
\end{equation}
As in the fermion case we can still make UV-independent conclusions regarding the model since the strongest bounds on the parameter space for $\mu_0 > m_Z$ come from indirect detection.

\begin{figure}
    \centering
    \includegraphics[width=0.8\textwidth]{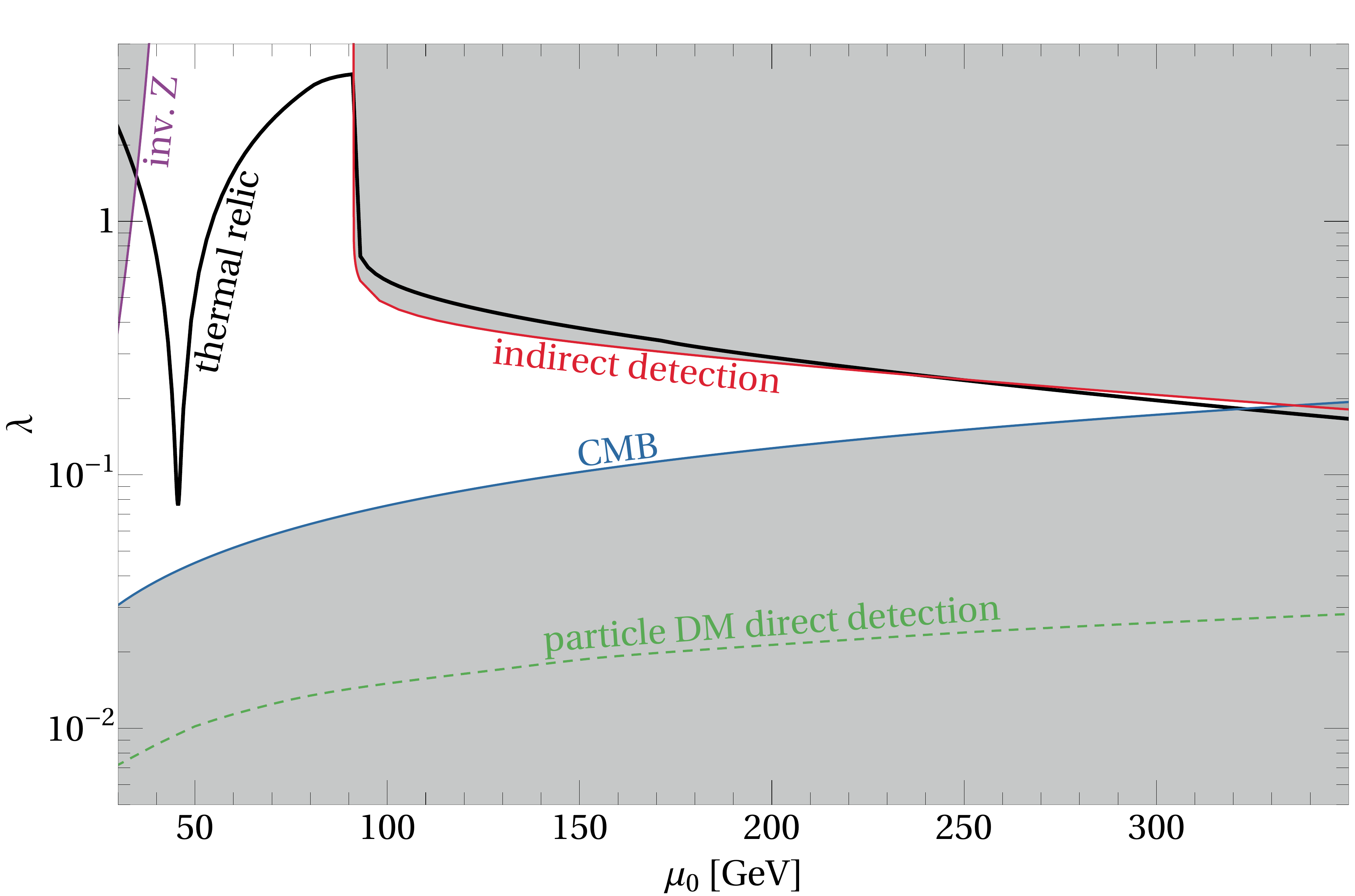}
    \caption{Parameter space for continuum vector $Z$-portal DM. The black curve reproduces the observed thermal relic density~\cite{Planck:2015fie}. We show experimental constraints from CMB observations \textit{(blue)}~\cite{Slatyer:2016qyl,Slatyer:2015jla}, indirect detection with the Fermi-LAT experiment \textit{(red)}~\cite{Fermi-LAT:2016uux,Abazajian:2020tww}, and the LEP bound on the invisible $Z$ width \textit{(purple)}~\cite{ALEPH:2005ab}. We also show the LUX-ZEPLIN direct detection bound for ordinary particle vector $Z$-portal DM \textit{(green, dashed)}~\cite{LZ:2022ufs}.}
    \label{fig:vectorbounds}
\end{figure}

In fig.~\ref{fig:vectorbounds} we show the region of parameter space that yields the observed dark matter relic density, alongside current experimental bounds and the direct detection bound for ordinary particle DM. Continuum abelian vector $Z$-portal DM reproduces the correct relic abundance and is compatible with experimental constraints for gap scales in the range $\mu_0 \in (35, 90)$~GeV. One might note another narrow window of phenomenologically viable parameter space around $250$~GeV~$< \mu_0 < 300$~GeV in fig.~\ref{fig:vectorbounds}. This would probably be ruled out by an update of the indirect detection bound, however. We will comment further on this in our conclusions.

%%%%%%%%%%%%%%%%%%%%%%%%%%%%%%%%%%%%%%%%%%%%%%%%%%%%%%%%%%%%%%%%%%%
%%%%%%%%%%%%%%%%%%%%%%%%%%%%%%%%%%%%%%%%%%%%%%%%%%%%%%%%%%%%%%%%%%%
%%%%%%%%%%%%%%%%%%%%%%%%%%%%%%%%%%%%%%%%%%%%%%%%%%%%%%%%%%%%%%%%%%%
%%%%%%%%%%%%%%%%%%%%%%%%%%%%%%%%%%%%%%%%%%%%%%%%%%%%%%%%%%%%%%%%%%%

\subsection{Experimental bounds}
The bounds on the parameter space follow analogously to the fermion model. Indirect detection is only relevant for $\mu_0 > m_Z$. We compute the $ZZ$ cross section and compare to the Fermi-LAT upper limit~\cite{Fermi-LAT:2016uux,Abazajian:2020tww}. This excludes the model for $m_Z < \mu_0 < 250$~GeV, as depicted in fig.~\ref{fig:vectorbounds}.

We again take $\rho_0 = 2\pi$ for the remaining constraints. The decay width of DM states at late times is the same as for scalar continuum DM, i.e. eq.~\eqref{eq:latedecaywidth} times $2/3$, with the replacement $\sin^2 \alpha \rightarrow \lambda$. The resulting bound from reionization of hydrogen after recombination rules out gap scales larger than $300$~GeV.

Below about $\mu_0 = 35$~GeV, the model is excluded by invisible $Z$ decays, $Z \rightarrow VV$. The decay width is given by
\begin{equation}\begin{split}
    &\left(\frac{\rho_0}{2\pi}\right)^2 \frac{g_Z^2 \lambda^2}{768\pi} m_Z \int d(t_1^2) \int d(t_2^2) \sqrt{t_1^2-1} \sqrt{t_2^2-1}  \frac{\sqrt{1 - 2x(t_1^2+t_2^2) + x^2(t_1^2-t_2^2)^2}}{x^2 t_1^2 t_2^2}  \\
    &\times  \left[x(t_1^2+t_2^2) -2x^2(t_1^4 + 6t_1^2t_2^2 + t_2^4) + x^3(t_1^2+t_2^2)(t_1^4 + 14t_1^2t_2^2 +t_2^4) + 4x^4 t_1^2 t_2^2 (t_1^2-t_2^2)^2 \right]
\end{split}\end{equation}
with $x$, $t_1$, $t_2$ and the integration region defined as in eq.~\eqref{eq:fermionZdecay}, and we again use the LEP bound $\Gamma_{\rm inv} < 2$~MeV~\cite{ALEPH:2005ab}.

We should also consider invisible decays of the Higgs $h \rightarrow VV$, which could arise from a term in the Lagrangian like $V_\mu V^\mu H^\dagger H$. Such a term is generated by mixing of the dark Higgs (which was needed to lift the zero mode to mass $M$) with the SM Higgs. Even if there is no mixing at tree level, it will be induced by loop diagrams involving the fermions in eq.~\eqref{eq:weylfermions}. We therefore expect the $h \rightarrow VV$ decay width to be suppressed by $(v/M)^2$. We assume that $M$ is sufficiently large to suppress the decay width below the LHC upper limit on Higgs invisible decays~\cite{CMS:2022qva,ATLAS:2022yvh}, so we do not plot any corresponding bound in fig.~\ref{fig:vectorbounds}.

Lastly, we consider direct detection. The spin-dependent DM-nucleon cross section in the nonrelativistic limit is
\begin{equation}
    \int \frac{d \mu'^2}{2\pi} \rho(\mu'^2) \frac{g_Z^4 \lambda^2 \alpha \mu_{\psi N}^2 f_N^2}{2 \pi m_Z^4} ,
\end{equation}
which is the same as the fermion model (eq.~\eqref{eq:DDcrosssection}) up to a factor of $2/3$ and with the replacement $\sin^2 \alpha \rightarrow \lambda$. We compute the direct detection constraint for ordinary particle DM by ignoring continuum suppression and comparing to the LUX-ZEPLIN limit on the DM-neutron spin-dependent cross section~\cite{LZ:2022ufs}. It is clear from fig.~\ref{fig:vectorbounds} that this model would be completely ruled out over our entire range of parameter space, were the DM an ordinary particle.

Continuum kinematics suppresses direct detection cross sections by three to four orders of magnitude, similar to the fermion model. Thus, in sharp contrast to particle DM, direct detection does not constrain our continuum model at all. For gap scales in between $35$~GeV and $m_Z \sim 90$~GeV we reproduce the observed relic density while being consistent with all experimental bounds.

%%%%%%%%%%%%%%%%%%%%%%%%%%%%%%%%%%%%%%%%%%%%%%%%%%%%%%%%%%%
%%%%%%%%%%%%%%%%%%%%%%%%%%%%%%%%%%%%%%%%%%%%%%%%%%%%%%%%%%%
\section{Conclusions and outlook}\label{sec:conclusions}
%%%%%%%%%%%%%%%%%%%%%%%%%%%%%%%%%%%%%%%%%%%%%%%%%%%%%%%%%%%
%%%%%%%%%%%%%%%%%%%%%%%%%%%%%%%%%%%%%%%%%%%%%%%%%%%%%%%%%%%

In this paper we have constructed models of fermion and vector continuum DM, annihilating through the $Z$ portal. We found that these simple models are compatible with experimental constraints for gap scales $\mu_0 \in (60, 200)$~GeV and $\mu_0 \in (35, 90)$~GeV for the fermion and vector cases, respectively. Owing to the strong continuum kinematic suppression of DM-nucleon scattering cross sections, current direct detection bounds do not constrain our model. Minimal $Z$-portal DM models with ordinary particles, on the other hand, are essentially excluded by direct detection experiments.

The main constraints on our models arise instead from reionization of hydrogen after recombination (which is unique to continuum DM), invisible decays of the $Z$ and Higgs, and indirect detection. It is worth noting that the most recent indirect detection bounds published by the Fermi-LAT collaboration are six years old~\cite{Fermi-LAT:2016uux}; the bound on the $ZZ$ cross section we used~\cite{Abazajian:2020tww} does not incorporate newer data than this, and the bound we quoted for the $Zh$ cross section~\cite{Arcadi:2014lta} is even older. It would be interesting to perform an updated analysis with the full fourteen years of Fermi data (2008--2022), along the lines of ref.~\cite{Dessert:2022evk}, in the context of general portal models. One could potentially probe the fermion model for $\mu_0 > m_Z$ in this way. Such an analysis would also probably exclude the small window between $\mu_0 = 250$~GeV and $300$~GeV currently permitted for the vector model (see fig.~\ref{fig:vectorbounds}).

Furthermore, we mentioned briefly that continuum states can undergo cascade decays, which may lead to distinctive collider signatures. A dedicated study of continuum collider signatures would be a promising avenue for further probing continuum DM models. This is especially important for studying the parameter space at gap scales less than $m_Z$, where neither indirect nor direct detection experiments are especially sensitive to our models.

We have remained agnostic about the microscopic origin of the continuum where possible. Warped 5D constructions of gapped continua necessarily lead to additional gravitational modes in the 4D effective theory, namely Kaluza--Klein gravitons and the radion, the effects of which we have ignored. More generally, consistently coupling a continuum to gravity requires a nonstandard gravitational sector~\cite{Fichet:2022ixi}. It would be interesting to study the interplay of the gravitational sector with continuum DM, and whether this might lead to new experimental signatures.

The results of this work, together with refs.~\cite{Csaki:2021gfm,Csaki:2021xpy}, establish continuum $Z$-portal models for scalar, fermion, and vector DM --- all of which are phenomenologically viable. This is only the tip of the iceberg for continuum DM model building. Our continuum fermion model is most similar to Majorana fermion particle DM, so one could consider a Dirac fermion instead. The main obstacle to building such a model is in constructing a gapped continuum for the Dirac fermion. 5D constructions generally lead to chiral fermions in the 4D effective theory; obtaining a continuum Dirac fermion would require two bulk fermions in the 5D theory coupled to each other~\cite{Contino:2004vy,Cacciapaglia:2008ns}. For some work along these lines see refs.~\cite{Jacobs:2014nia,Liu:2018bye,Plantz:2018tqf}; also see refs.~\cite{Lee:2008xf,Liu:2009dm,Cubrovic:2009ye} for discussion of continuum Dirac fermions in (2+1) dimensions. Likewise, we have only considered abelian vector DM here, so it may be interesting to study the nonabelian case. Effective $Z$-portal interactions for nonabelian gauge bosons have been written down in the literature~\cite{Arcadi:2017kky,Ellis:2017ndg,Escudero:2016gzx}, but it is difficult to construct a theory to generate these interactions.

Another option is to consider different portal interactions. A minimal Higgs portal model does not work for continuum DM because the continuum decay width is suppressed by $(m_f/m_h)^2$, where $m_f$ is the mass of the outgoing fermion, which is incompatible with CMB constraints. However, a new physics portal such as a $Z'$ could work, or one could construct a nonminimal Higgs portal where the continuum DM decays to a different dark sector particle.

Going beyond the WIMP thermal freeze-out paradigm, there are myriad opportunities for continuum DM model building. Recalling that extra-dimensional constructions require new gravitational modes, it is natural to wonder if a continuum graviton, stabilized by KK-parity and with the relic abundance set by e.g. the freeze-in mechanism~\cite{Hall:2009bx,Elahi:2014fsa}, could be a viable DM candidate. Furthermore, any continuum state can generically have very weak couplings to the SM --- which is typically required for freeze-in DM --- if the cutoff scale is much larger than the gap scale.

In conclusion, continuum models are an exciting new direction in dark matter physics, with unique experimental signatures that differ from ordinary particle dark matter. We hope that the simple $Z$-portal models we have constructed here will inspire further continuum model building in the near future.

\acknowledgments 
We thank Steven Ferrante, Gowri Kurup, and Maxim Perelstein for helpful discussions and notes. CC and AI are supported in part by the NSF grant PHY-2014071. CC is also supported in part by the BSF grant 2020220. AI is also supported in part by NSERC, funding reference number 557763. The research activities of SL are supported by the Samsung Science Technology Foundation under Project Number SSTF-BA2201-06.

\appendix
%%%%%%%%%%%%%%%%%%%%%%%%%%%%%%%%%%%%%%%%%%%%%%%%%%%%%%%%%%%
%%%%%%%%%%%%%%%%%%%%%%%%%%%%%%%%%%%%%%%%%%%%%%%%%%%%%%%%%%%
\section{Gapped continua from a warped extra dimension}\label{sec:appendix}
%%%%%%%%%%%%%%%%%%%%%%%%%%%%%%%%%%%%%%%%%%%%%%%%%%%%%%%%%%%
%%%%%%%%%%%%%%%%%%%%%%%%%%%%%%%%%%%%%%%%%%%%%%%%%%%%%%%%%%%

In this appendix we will briefly review the derivation of 4D gapped continua from 5D constructions~\cite{Csaki:2021gfm,Megias:2019vdb}, then prove the spectral density obeys eq.~\eqref{eq:universalspectraldensity} near the gap scale.

\subsection{Setup}
We consider a 5D spacetime with metric
\begin{equation}
    ds^2 = e^{-2A(z)} \left( \eta_{\mu\nu} dx^\mu dx^\nu - dz^2 \right)
\end{equation}
where the conformal coordinate $z$ ranges from $R$, the location of the UV brane, to $\infty$. We will assume the warp factor satisfies $A''(z) < 0$. As a simple example, consider a scalar field $\Phi(x, z)$ which propagates in the 5D bulk, with action
\begin{equation}\label{eq:scalaraction}
    \frac{1}{2} \int d^4 x \int dz \sqrt{g} \left[ \left( D_a \Phi \right)^2 - m^2 \Phi^2 \right]
\end{equation}
(we use Latin indices for 5D coordinates and Greek indices for 4D coordinates).

To compute the boundary effective action we need to integrate out the extra dimension. We rescale the field $\Phi \rightarrow \sqrt{R} e^{-3A/2} \Phi$, then decompose the field in 4D momentum space as
\begin{equation}\label{eq:decomposition}
    \Phi(p, z) = \phi(p) f(p, z) .
\end{equation}
The wavefunctions $f(\mu, z)$ are determined by solving the classical 5D equation of motion. In fact, this takes the form of a Schr\"odinger-like equation:
\begin{equation}
    -f''(\mu,z) + V(z) f(\mu,z) = \mu^2 f(\mu,z)
\end{equation}
where the potential is
\begin{equation}
    V(z) = \frac{9}{4} A'(z)^2 - \frac{3}{2} A''(z) + m^2 e^{-2A(z)} .
\end{equation}
A continuum with gap scale $\mu_0$ exists if $V \sim \mu_0^2$ as $z \rightarrow \infty$. It is convenient to shift the potential as $V \rightarrow V - \mu_0^2$, so that the Schr\"odinger equation becomes
\begin{equation}\label{eq:schrodinger}
    -f''(\mu,z) + V(z) f(\mu,z) = \kappa^2 f(\mu,z)
\end{equation}
where $\kappa^2 = \mu^2 - \mu_0^2$, and $V$ tends to zero at large $z$.

We can integrate the action, eq.~\eqref{eq:scalaraction}, by parts. This leads to a bulk term that vanishes on the equation of motion, and a nonvanishing boundary term which corresponds to the boundary effective action:
\begin{equation}
    \int d^4 x \frac{1}{2R} e^{-3A/2} \Phi \partial_z \left( e^{3A/2} \Phi \right) \Big |_{z=R} .
\end{equation}
Rewriting the effective action in momentum space and using eq.~\eqref{eq:decomposition}, we find
\begin{equation}\begin{split}
    \frac{1}{2R} &\int \frac{d^4 p}{(2\pi)^4} e^{-3A/2} \phi(p) f(p, z) \partial_z \left( e^{3A/2} f(p,z) \right) \phi(p) \Big |_{z=R} \\
    = \frac{1}{2R} &\int \frac{d^4 p}{(2\pi)^4} \Phi(p, R) \left[ 3A'/2 + \frac{f'(p,z)}{f(p,z)} \right] \Big |_{z=R} \Phi(p, R)  .
\end{split}\end{equation}
We can then read off the spectral density function:
\begin{equation}
    \rho(\mu^2) = -2 R \Im \left[ \left(\frac{3A'}{2} + \frac{f'(\mu,z)}{f(\mu,z)} \right)^{-1} \Big |_{z=R} \right] .
\end{equation}
We have glossed over some subtleties concerning the normalization of the spectral density which are unimportant for the present purpose.

Note that the natural size of the spectral density is set by $1/R^2$. For fermions and vector bosons (as well as other particles), the spectral density is still determined by wavefunctions and their derivatives evaluated at $z = R$. The wavefunctions satisfy the Schr\"odinger-like equation~\eqref{eq:schrodinger}, but with a different potential. In what follows we will simply quote expressions for the potential and the spectral densities, and refer the reader to ref.~\cite{Megias:2019vdb} for details of their derivation.

\subsection{Spectral density near the gap scale}
We now show that the spectral density near the gap scale takes the universal form
\begin{equation}
    \rho(\mu^2) = \frac{\rho_0}{\mu_0^2} \sqrt{\frac{\mu^2}{\mu_0^2} - 1}
\end{equation}
where $\rho_0$ is a constant. Ref.~\cite{Csaki:2021gfm} proved this for scalars; we would like to extend the argument to spin-$1/2$ and spin-$1$ particles. We begin with the spin-$1$ case, which closely follows the proof for scalars.

\subsubsection{Vector bosons}
For vector bosons, the spectral density is given by
\begin{equation}\label{eq:vectorspectraldensity}
    \rho(\mu^2) = -2R \Im \left( \frac{f(\mu, z)}{f'(\mu,z)} \right) \Big |_{z=R}
\end{equation}
where the wavefunctions $f$ satisfy eq.~\eqref{eq:schrodinger} with potential
\begin{equation}
    V(z) = \frac{1}{4}A'(z)^2 - \frac{1}{2}A''(z) - \mu_0^2 .
\end{equation}
Recall we have shifted the potential so that as $z \rightarrow \infty$, $V \rightarrow 0$. For finite $z$, $V$ is always positive due to our assumption $A'' < 0$.
We can rewrite eq.~\eqref{eq:vectorspectraldensity} as
\begin{equation}\label{eq:vectorspectraldensity2}
    \rho(\mu^2) = 2 R \left\lvert \frac{f(\mu, R)}{f'(\mu,R)} \right\rvert^{2} \Im \left( \frac{d \log f(\mu, z) }{dz} \right) \Big |_{z=R} ,
\end{equation}
which indicates the behavior of the spectral density near the gap scale is determined by the $\Im d \log f(\mu,z) / dz |_{z=R}$ term.

Now, the Schr\"odinger-like equation in general has two real solutions, $f_1$ and $f_2$, so the general solution is
\begin{equation}
    f = f_1 + c f_2 ,
\end{equation}
where $c$ is a complex constant determined by boundary conditions. At large $z$ we can solve the equation since $V \rightarrow 0$ to get $f_1 = \cos \kappa z$, $f_2 = \sin \kappa z$. Applying the outgoing wave boundary condition, $f \sim e^{i\kappa z}$, determines $c = i$. We can then evalute
\begin{equation}
    \Im \left( \frac{d \log f(\mu, z) }{dz} \right) \Big |_{z=R} = \frac{f_1 f_2' - f_1' f_2}{\lvert f_1 \rvert^2 + \lvert f_2 \rvert^2} \Big |_{z=R} .
\end{equation}
The quantity in the numerator is the Wronskian, which is independent of $z$ because the Schr\"odinger equation has no first-derivative term (this is a consequence of Abel's identity). We can therefore evaluate it at large $z$, where we know $f_1$ and $f_2$, to find
\begin{equation}
    f_1 f_2' - f_1' f_2 = \kappa .
\end{equation}
Under the assumption $\lvert f(\mu,R) \rvert^2 \neq 0$ (to be justified shortly), we then have
\begin{equation}
    \Im \left( \frac{d \log f(\mu, z) }{dz} \right) \Big |_{z=R} \propto \kappa .
\end{equation}
Comparing to eq.~\eqref{eq:vectorspectraldensity2}, we see that if we further assume $f'(\mu_0,z)/f(\mu_0, z)$ does not vanish or blow up at $z = R$, $\rho(\mu^2) \propto \kappa = \sqrt{\mu^2 - \mu_0^2}$ near the gap scale.

To finish up the proof, we need to justify the two assumptions we made. To do this, we first multiply the Schr\"odinger equation (eq.~\eqref{eq:schrodinger}) by $f^*$ and integrate it from $z$ to $\infty$. After an integration by parts and evaluating near the gap scale, it is easy to see
\begin{equation}
    \frac{d}{dz} \lvert f(\mu, z) \rvert^2 < 0
\end{equation}
for all $z$. We know that $\lim_{z \rightarrow \infty} \lvert f(\mu,z) \rvert^2 = 1$, and therefore 
\begin{equation}
    \lvert f(\mu, z) \rvert^2 > 1
\end{equation}
for all $z$, justifying our assumption $\lvert f(\mu,R) \rvert^2 \neq 0$. It also trivially follows that $f'(\mu_0,z)/f(\mu_0,z)$ does not vanish. Lastly, we will show by contradiction that $f'(\mu_0,z)/f(\mu_0,z)$ does not exhibit singular behaviour. We rewrite the Schr\"odinger equation (with $\kappa = 0$) as
\begin{equation}
    \left( \frac{f'}{f} \right)' + \left( \frac{f'}{f} \right)^2 = V(z) .
\end{equation}
Suppose $f'/f$ did in fact blow up at some point $z_0$. Then in the vicinity of $z_0$, we can approximate the potential by the positive constant $V(z_0) \equiv V_0$. The Schr\"odinger equation can then be solved near $z_0$, and we find
\begin{equation}
    \frac{f'}{f} \sim \sqrt{V_0} \tanh \left( \sqrt{V_0} (z - z_0) \right) .
\end{equation}
But this does not exhibit singular behaviour at $z_0$, so we have reached a contradiction. In fact, by a similar argument one can show $f'(\mu_0, z)/f(\mu_0,z)$ can only blow up when $V(z) = 0$. This completes the proof.

\subsubsection{Fermions}
We consider left-handed continuum fermions here, with the argument easily extended to right-handed fermions. The spectral density is
\begin{equation}\label{eq:fermionspectraldensity}
    \rho(\mu^2) = -\frac{2R}{\mu} \Im \frac{f^L(\mu,z)}{f^R(\mu,z)} \Big |_{z=R} = -\frac{2R}{\mu} \frac{1}{\lvert f^R \rvert^2} \Im f^L f^{R*} \Big |_{z=R} ,
\end{equation}
where $f^L$ and $f^R$ are wavefunctions for left-handed and right-handed fermions. They each satisfy eq.~\eqref{eq:schrodinger}, but with different potentials:
\begin{equation}
    V^{L,R} = e^{-2A(z)} M(z)^2 \pm e^{-A(z)} \left( M'(z) - M(z) A'(z) \right) -\mu_0^2 .
\end{equation}
In this equation $M(z)$ is a $z$-dependent bulk mass term, which would arise from a Yukawa interaction with a scalar field that has a $z$-dependent vev. We assume $M(z) \sim \mu_0 e^{A}$ as $z \rightarrow \infty$. Then the $V^{L,R}$ approach 0 at large $z$. We will also assume the potentials are positive for any finite $z$.

The Schr\"odinger equation again admits two real solutions for each of $f^L$ and $f^R$, and from the arguments for the vector boson case we know the outgoing wave boundary condition fixes
\begin{equation}
    f^{L,R} = f^{L,R}_1 + i f^{L,R}_2 .
\end{equation}
Note that in general $f^L_i \neq f^R_i$. We then have
\begin{equation}\label{eq:fermionimaginarypart}
    -\Im f^L f^{R*} = f^L_2 f^R_2 \left[ \frac{f^L_1}{f^L_2} - \frac{f^R_1}{f^R_2} \right] .
\end{equation}
But the derivative of the term in brackets is proportional to the Wronskian, which we know equals $\kappa$ for both $f^L$ and $f^R$:
\begin{equation}
    \frac{d}{dz}\left[ \frac{f^L_1}{f^L_2} - \frac{f^R_1}{f^R_2} \right] = \frac{f^{L'}_1 f^L_2 - f^L_1 f^{L'}_2}{(f^L_2)^2} - ({\rm L} \leftrightarrow {\rm R}) = \kappa \left( \frac{1}{(f^R_2)^2} - \frac{1}{(f^L_2)^2} \right) .
\end{equation}
Integrating this equation from $z = \infty$ to $z = R$, and substituting back into eq.~\eqref{eq:fermionimaginarypart}, we find
\begin{equation}
    -\Im f^L f^{R*} \Big |_{z=R} = \kappa f^L_2(R) f^R_2(R) \int_{R}^\infty dz \left( \frac{1}{(f^L_2)^2} - \frac{1}{(f^R_2)^2} \right) .
\end{equation}
It follows that the spectral density near the gap scale is proportional to $\kappa$, as long as $f^{L,R}_2$ are nonvanishing at $z=R$, and the integral is nonvanishing and finite. We will not try to prove these requirements. However, one can check the simple case where the $V_{L,R}$ fall off exponentially at large $z$. Then an explicit form for the wavefunctions can be given in terms of Bessel functions, and one finds the spectral density is indeed proportional to $\kappa$ near the gap scale.

Admittedly, the argument for fermions is less elegant than that for scalars or vectors. It would be interesting to pursue a more concise argument that required fewer assumptions about the behavior of the wavefunctions at $z=R$.

\bibliographystyle{JHEP}
\bibliography{ContinuumDMv3}
\end{document}